# Geodesic conformal transformation optics: manipulating light with continuous refractive index profile

Lin Xu[1], Tomáš Tyc[2] and Huanyang Chen[1]*

[1] Institute of Electromagnetics and Acoustics and Department of Electronic Science, Xiamen University Xiamen 361005, China

[2] Department of Theoretical Physics and Astrophysics, Masaryk University, Kotlarska 2, 61137 Brno, Czech Republic

**Conformal transformation optics provides a simple scheme for manipulating light rays with inhomogeneous isotropic dielectrics. However, there is usually discontinuity for refractive index profile at branch cuts of different virtual Riemann sheets, hence compromising the functionalities. To deal with that, we present a special method for conformal transformation optics based on the concept of geodesic lens. The requirement is a continuous refractive index profile of dielectrics, which shows almost perfect performance of designed devices. We demonstrate such a proposal by achieving conformal transparency and reflection. We can further achieve conformal invisible cloaks by two techniques with perfect electromagnetic conductors. The geodesic concept may also find applications in other waves that obey the Helmholtz equation in two dimensions.**

*Introduction.-*Based on covariance of Maxwell's equations and multi-linear constitutive equations, optical property of virtual space and physical space could be connected by a coordinate mapping [1]. In 2006, Leonhardt [2] presented that a conformal coordinate mapping between two complex planes could be performed for scalar field of refractive index of dielectrics such that light rays could be manipulated freely. Coincidentally, Pendry et al [3] provided a general method for controlling electromagnetic field in space of three dimensions. These two seminal papers launched a new research field named transformation optics (TO) [4-7], which originally mainly focused on optical invisibility. After that, a lot of optical designs based on TO popped up, such as carpet cloak[8], illusion optics devices[9], field rotator[10] and so on. With the development of metamaterials, several proof-of-principle experiments have been achieved, like reduced cloak in two dimensions [11], carpet cloaks [12, 13], field rotator [14] and concentrator [14]. Besides metamaterials, structured materials like photonic crystals [15] and waveguides [16, 17], could also serve for designing transformation materials, which provide different platforms. In addition, the idea of TO has also been extended to other waves, such as acoustics [18], plasmonics [19] and thermodynamics [20-22].

From geometrical perspective, TO has been established for the connection between curved space-time and multi-linear response of structured materials, which gives a blueprint to control light propagation. Despite these versatile design proposals and proof-of-principle experiments, TO still encounter challenges in practical engineering because of the complexity of the required materials. Take the design of invisibility for example, the required tensor fields of permittivity and permeability have infinity values at some points [3]. At the same time, conformal transformation optics (CTO) (in two-dimensional space) was proposed by Leonhardt [2] as a simpler scheme for manipulating light rays by using dielectrics with inhomogeneous isotropic refractive index profile, which is a scalar field. After that, lots of designs with CTO have been proposed, like conformal invisible cloaks [23-25], conformal transparency devices [26], conformal illusion devices [27] and conformal Talbot devices [17]. Though some experiments [17, 28, 29] have also demonstrated the principle of CTO [30], there is still a need to optimize the design procedure, so that a feasible scalar field of refractive index can be obtained for practical applications. Moreover, in several proposals of CTO [2, 24, 25], there is discontinuity of refractive index profile, which might have some unwanted influence on the performance of CTO in turn.

In this letter, to solve the problems caused by discontinuity of refractive index profile, we propose a special kind of CTO based on the concept of geodesic lens [31, 32], which we would like to call "the geodesic conformal transformation optics" (GCTO). The mapping in GCTO is a composite mapping of geodesic conformal mapping and analytical conformal mapping, which could map an artificial Riemann surface (virtual space) with homogeneous refractive index profile to a plane (physical space) with inhomogeneous refractive index profile. The requirement of inhomogeneous refractive index is a continuous scalar field. We demonstrate our method by achieving optical transparency and wave reflection. To obtain invisible cloak, we develop two techniques with perfect electromagnetic conductors (PECs). Moreover, we explain that our method can work not only in the geometric-optical limit, but can also be extended to wave regime at a discrete series of frequencies.

Table 1 Refractive index profiles of three absolute instruments and spectrum of corresponding geodesic lens.

| Lens | Refractive index profile | Geodesic lens | Description of geodesic lens | Spectrum |
|---|---|---|---|---|
| Maxwell's fish-eye lens | $n(r) = \dfrac{2}{1+(r/r_0)^2}$ | Sphere | $h(\rho) = \arcsin(\rho)$ | $\dfrac{\omega r_0}{c} = \sqrt{(N+m)(N+m+1)}$ $\approx N+m+0.5,$ |
| Inverse invisible lens | $r > r_0, (r/r_0)n^{3/2} + (r/r_0)n^{1/2} - 2 = 0;$ $r < r_0, n = 1$ | Truncated Tannery's pear | $h_1(\rho) = -\rho + 2\arcsin(\rho)$ $h_2(\rho) = 2 + \pi - \rho$ | $\dfrac{\omega r_0}{c} \approx N+m+0.5,$ |
| Generalized Maxwell's fish-eye lens | $n(r) = \dfrac{2(r/r_0)^{1/M-1}}{1+(r/r_0)^{2/M}},$ M=2,3,4... | Spindle | $h(\rho) = \arcsin(M\rho)$ M=2,3,4... | $\dfrac{\omega r_0}{c} = \dfrac{\sqrt{(N+M\cdot m)(N+M\cdot m+1)}}{M}$ $\approx \dfrac{N+0.5}{M} + m,$ |

*This spectrum is numeric approximate by WKB method, not exact result [33, 34].

*Analytical conformal mapping and geodesic conformal mapping.* –In traditional conformal transformation optics, we usually use analytical function w=f(z) to connect complex plane z (physical space) with complex plane w (virtual space). This analytical conformal mapping (ACM) preserves the angle of two intersecting lines because of Cauchy-Riemann conditions [35]. The refractive index profiles of two complex planes have the relation of Eq. (1) based on CTO [2, 30].

$$n(z) = |dw/dz| n(w) \qquad (1)$$

In general, two-dimensional surfaces are all conformally flat and they differ by a scalar curvature field [36]. For light rays, such a curvature field could be treated as scalar field of refractive index profile. There is a kind of coordinate transformation from inhomogeneous plane lens with cylindrically-symmetric refractive index profile n(r) to homogeneous geodesic lens with refractive index profile n'(h)=1, which sets up equivalence of the corresponding optical path elements [31]. This mapping is called geodesic conformal mapping (GCM), written as

$$\rho = n(r)r, dh = n(r)dr, \qquad (2)$$

where ρ is radial coordinate, and h(ρ) is the length measured along the meridian from North pole in the geodesic surface, as shown with examples in Fig. 1. One important property of GCM is that it also preserves angle of two intersection lines, where its name comes from. Original GCM was used to deal with three-dimensional Luneburg inverse problem associated with spherical media [31, 32]. In this letter, we will focus on two-dimensional conformal transformation. Therefore, GCM here is used in a simpler version.

For the sake of later discussion, we list three typical geodesic lenses (sphere, truncated Tannery's pear and spindle) and the corresponding inhomogeneous plane lenses (Maxwell's fish-eye lens, inverse invisible lens and Generalized Maxwell's fish-eye lens) with their properties summarized in Table 1 (for details, see in Ref. [34]). Tannery's pear is a two-dimensional compact surface and its geodesics are closed [37]. It turns out that geodesics of truncated Tannery's pear are also closed [32, 34]. Spindle is equivalent to a portion of 1/M of sphere, which is glued along its boundary [31, 32, 34, 38]. All light rays form closed trajectories in these geodesic lenses and the corresponding plane lenses as shown in Fig. 1. Moreover, these three plane lenses are also called absolute instruments, where closed light rays and perfect imaging could be achieved [31, 39]. We will use these geodesic lenses for further work in the following sections. In wave regime, these geodesic lenses and plane lenses have the spectrum shown in the fifth column of Table 1. It turns out that their spectrum are highly degenerated [34]. This is vital for the wave performance of our design.

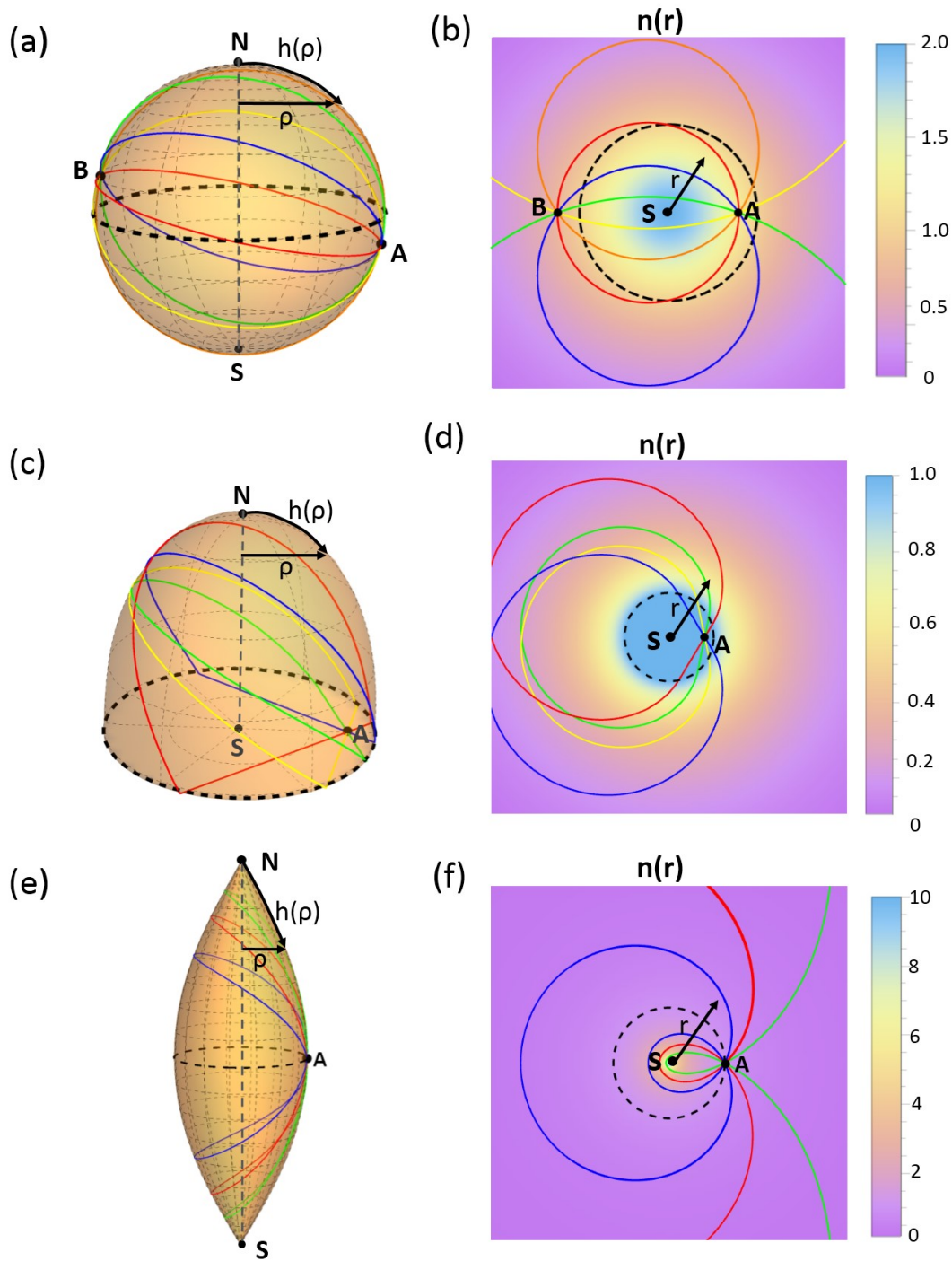

Fig.1 Geodesic lenses (left column) and corresponding plane lens with rotational symmetry (right column). In geodesic lenses, the rotational symmetry axis is the line which connects North (N) pole and South (S) pole. $\rho$ is the radial coordinate, and $h(\rho)$ is the length measured along the meridian from N pole in the geodesic surface. Light rays starting from point A form closed trajectories shown in different colors. Those geodesic lenses are sphere (a), truncated Tannery's pear (c) and spindle (e). Their descriptions are shown in Table 1. Their corresponding inhomogeneous plane lenses are Maxwell's fish-eye lens (b), inverse

invisible lens (d) and generalized Maxwell's fish-eye (f), respectively. Contour plots show refractive index profile of n(r) with cylindrical symmetry. Point S represents the origin of the plane, which mapped from S pole of geodesic lenses. The infinity corresponds to N pole. Colored light rays are mapped from those of geodesic lenses. Dashed black lines are places with refractive index of unity at radius of $r_0$.

*Geodesic conformal transformations optics.* – Transformation optics show us a heuristic method to understand certain complex medium as curved space for light [36]. Therefore, we usually start from a simple space, namely flat Euclidean space, to design transformation medium to achieve a certain property, like invisible cloak [3]. Sometimes, we could also start from a non-Euclidean space to make an optical design with a fancy functionality [40]. Here, by compositing geodesic conformal mapping and analytical conformal mapping, we could achieve a mapping from virtual space that is homogeneous but not globally flat to a plane physical space with non-trivial metric. Using composite conformal mapping, we develop a special conformal method called geodesic conformal transformation optics to design optical devices in two dimensions. It gives geometric intuition for us to deal with problems in virtual space embedding in three dimensions, rather than in physical space plane with complex scalar field.

Let us start from a concrete example of conformal transparency to illustrate our method. Suppose that we have a virtual space with a sphere and a meshed plane in Fig. 2(a). The radius of sphere is R. They are connected by a branch cut (a solid arc in purple). For light rays in plane, they travel along straight lines. When parallel light rays shown in different colors propagate in plane and meet the branch cut, they enter a sphere and form closed trajectories of great circles, which are geodesics on the sphere. After they come back to the branch cut again on the sphere, they return to the plane to continue their journey with their positions and directions preserved. Compared with parallel light rays not entering into sphere, they have additional optical path on the sphere. But for observer far away, this sphere appears invisible in geometrical limit.

After introducing a virtual space in Fig. 2(a), we perform a geodesic conformal mapping (written as Eq. (2)) of the sphere to obtain a Maxwell's fish-eye lens shown in vertical plane in Fig. 2(b), which still connect original plane along the branch cut. With a further exponential conformal mapping, namely $w_2= \exp(w_1)$ of Maxwell's fish-eye lens, we obtain a cylindrical Mikealian lens with a period of $8\pi$ shown in Fig. 2(c) [41]. Finally, we could map the whole space of Fig. 2(c) to a physical space by an analytical conformal mapping, $w_1=z+4\log(z-a)-4\log(z+a)$, which results in a continuous refractive index profile shown in contour plot of Fig. 2(d). The analytical conformal mapping is carefully chosen such that branch cut is a quarter of a great circle in virtual space. Correspondingly light rays are shown in colored curves in Fig. 2(a-d). So far, we have achieved a conformal optical transparency by GCTO, which is a composition of one geodesic conformal mapping and two analytical conformal mappings. However, these mappings are carefully chosen such that virtual space and physical space are matched very well. Note that this profile is the same to that in Ref. [26], however, the current interpretation is more intrinsic hence will induce several designs in the following sections.

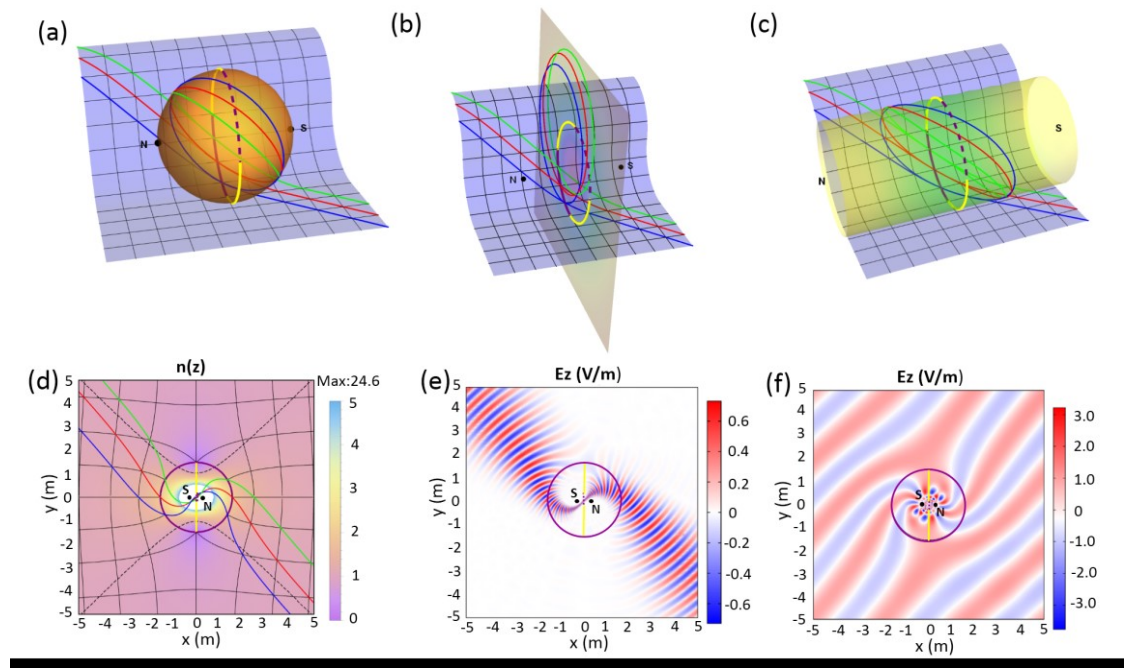

Fig. 2 Conformal transparency design with GCTO. Virtual space (a) consists of a complex plane and a sphere. They are sewed along a branch cut (solid arc in purple), which belongs to geodesics of both parts, namely a straight line in plane and a great circle in sphere. Dashed arc in purple is the image of the branch cut. The colored light rays are paralleled in plane and form closed trajectories in sphere. Light rays started from the branch cut cannot reach two yellow arcs. N and S denote the poles of the sphere. Using geodesic conformal mapping, sphere in (a) is mapped to Maxwell's fish-eye lens in (b). The corresponding light rays also form closed trajectories. With a further exponential conformal mapping, Maxwell's fish-eye lens in (b) is mapped to a cylindrical Mikealian lens shown in (c). Therefore by using an analytical conformal mapping, final physical space is obtained in (d), with contour plot of refractive index profile (dashed black lines are places with refractive index of unity). Corresponding light rays are shown in colored curves in (a-d). Gaussian beam in (e) and plane wave in (f) impinge our design at the angle of –π/4 at an eigen-frequency corresponding to N+m = 40 and N+m =10, respectively. Here we set the radius of sphere to be R=2 and the length of branch cut to be a quarter of a great circle for the illustration. The corresponding parameter *a* in an analytical conformal mapping is 0.3125.

*Performance in wave-optical regime.* –It turns out that conformal transformation optics not only work in geometrical optics but also at eigen-frequencies in wave regime [30, 42]. Since GCTO connects the virtual space and physical space, we can analyze problem in simpler virtual space instead of physical space with complex medium. As shown in Table 1, the spectrum of sphere is $\frac{\omega r_0}{c} = \sqrt{(N+m)(N+m+1)} \approx N+m+0.5,$ where the number m is an integer that expresses dependence of the eigenmode on the azimuthal angle phi $\varphi$ as $\exp(im\varphi)$, and N is a non-negative integer expressing the number of nodes of the wave along the meridian between North and South poles [34]. It means that light wave impinges in virtual space forms standing wave at those frequencies. As has been shown previously for related devices[42-44], there is almost perfect invisibility at the resonant frequencies of the non-Euclidean part of virtual space. For our case this means illuminating the device by wave with frequency corresponding to an integer $l = N+m$. To test this with our design, we set $r_0$=2. Gaussian beam at eigen-frequency of N+m = 40 reveals the trajectories of light rays of our design in geometrical optics, see Fig. 2(e). In Fig. 2(f), we also use plane wave at eigen-frequency of N+m = 10 to show almost perfect transparency effect in wave regime. All numerical simulations in this letter are calculated by the commercial software COMSOL MULTIPHYSICS.

As shown above, light rays could be manipulated by continuous scalar field of refractive index profile in the method of GCTO. For waves, it turns out that the spectrum of geodesic lens with closed geodesics is highly degenerated [34]. Owing to perfect imaging of geodesic lens and their spectrum, linear superposition of individual point sources could recover after the same period of time [34]. These could explain almost perfect performance of conformal transparency at eigen-frequencies [26, 44]. The numerical simulations in Fig. 2 (e) and (f) also demonstrate such performance.

The main property of our design is that the resulting refractive index profile is continuous scalar field. We could carefully design virtual space to achieve more feasible parameters for GCTO. For the design of conformal transparency, we should take care of two things. One is that the branch cut should be geodesic of both surfaces in virtual space. Otherwise light rays would have a chance to cross the branch cut immediately once they have left it, which would compromise the effect of transparency. The other thing is that the length of branch cut should not be larger than half of a great circle. Otherwise some light rays entering the sphere through the branch cut could leave the sphere before completing the closed loop on it, which would also compromise the effect of transparency. Recently, by setting the length of branch cut half of the great circle, conformal transparency with an optimized refractive index profile ranging from 0 to 5.21 was achieved [26].

*Another design of conformal transparency.* -Suppose that we have a virtual space combined with a plane and a truncated Tannery's pear shown in Fig. 3(a), which are connected along a branch cut (a line in purple). The branch cut on both surfaces is part of geodesic which satisfies the requirement of GCTO. Parallel light rays shown in colored curves propagate in plane and meet the branch cut, they enter the truncated Tannery's pear and form closed trajectories which have been mentioned above. Then they come back to the branch cut and return into the plane to continue their journey with their positions and directions preserved. But for an observer far away from the truncated Tannery's pear, it appears invisible within geometrical optics.

Based on this virtual space, we could use a geodesic conformal mapping from the truncated Tannery's pear to a inverse invisible lens in Fig. 3(b). Then by further using Zhukowski conformal mapping, namely $w=z+1/z$ [2], the whole virtual space in Fig. 3(b) is mapped to a physical space in Fig. 3(c), which has a continuous refractive index profile ranging from 0 to 16.

This design of conformal transparency with GCTO is a composition of a geodesic conformal mapping and a Zhukowski analytical conformal mapping. Owing to that truncated Tannery's pear is also a geodesic lens, its spectrum is approximately $\frac{\omega r_0}{c} \approx N + m + 0.5$ from WKB method [33, 34], which is listed in Table 1. We also use numerical simulation to demonstrate the performance of our design with $r_0$=2, see Fig. 3(d) with Gaussian beam by setting N+m = 20 and Fig. 3(e) with plane wave by setting N+m = 6 to demonstrate the geometric-optical regime and wave-optical regime, respectively.

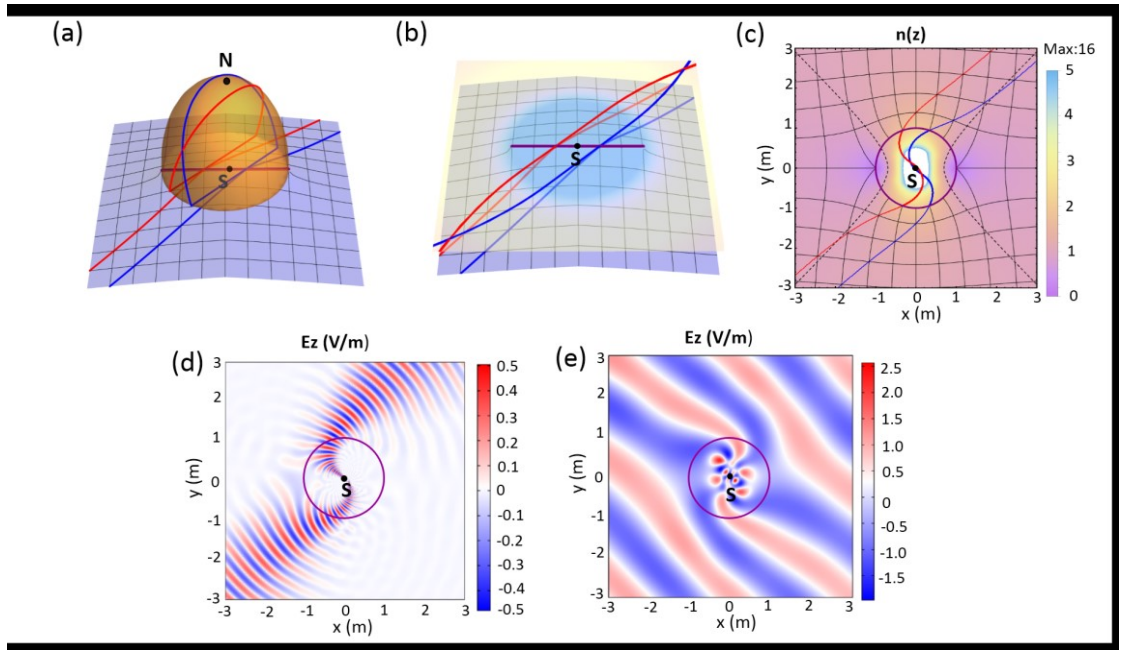

Fig. 3 Another conformal transparency design with GCTO. Virtual space (a) consists of a complex plane and truncated Tannery's pear. They are sewed along a branch cut (solid line in purple), which belongs to geodesics of both parts, namely a straight line in plane and a straight line of truncated Tannery's pear. Its length is 4. Two colored light rays (in red and in blue) are paralleled in plane and form closed trajectories in truncated Tannery's pear. Using geodesic conformal mapping, truncated Tannery's pear in (a) is mapped to inverse invisible lens in (b). Using a Zhukowski conformal mapping, (b) is further mapped to physical space in (c), where dashed black lines are places with refractive index of unity. Corresponding light rays (in red and in blue) are shown in colored curves in (a-d). Gaussian beam in (d) and plane

wave in (e) impinge our design at the angle of π/4 at eigen-frequency with N+m = 20 and N+m = 6, respectively. Here we set the length of branch cut to be 4 for the illustration.

*Wave Reflection-* Besides optical conformal transparency, we could also achieve wave omnidirectional reflection, which looks like a double-sided mirror from the outside. We start with virtual space, which is a combination of a plane and spindle surface shown in Fig. 4(a). They are sewed along a branch cut (an arc in purple). The branch cut is geodesic of both parts, namely it belongs to straight line of the plane and the equator of spindle surface. Parallel light rays shown in colored curves propagate initially in the plane. Those of them that cross the branch cut enter the spindle surface, travel along closed trajectories, and come back to the branch cut with their positions preserved but directions reflected. For observers far away from spindle surface, light rays seem to be totally reflected from the branch cut.

After performing geodesic conformal mapping and exponential analytical conformal mapping, the spindle surface is mapped to Mikealian lens with a period of 16π shown in Fig. 4(b). The period here is twice larger than that in conformal transparency design. By carefully choosing an analytical conformal mapping which is written as $w_1$=z+4log(z-*a*)-4log(z+*a*)), we map the whole structure to physical space in Fig. 4(c) with a contour plot of refractive index profile. The analytical conformal mapping here has slight difference from that in conformal transparency design. It can be even chosen such that branch cut could be the whole circle of spindle. In Fig. 4(d), we also perform numerical simulation to demonstrate wave reflection with Gaussian beam by setting N+m = 30. The conformal reflection works perfectly at any frequency, not just the resonant ones of conformal transparency in wave-optical regime.

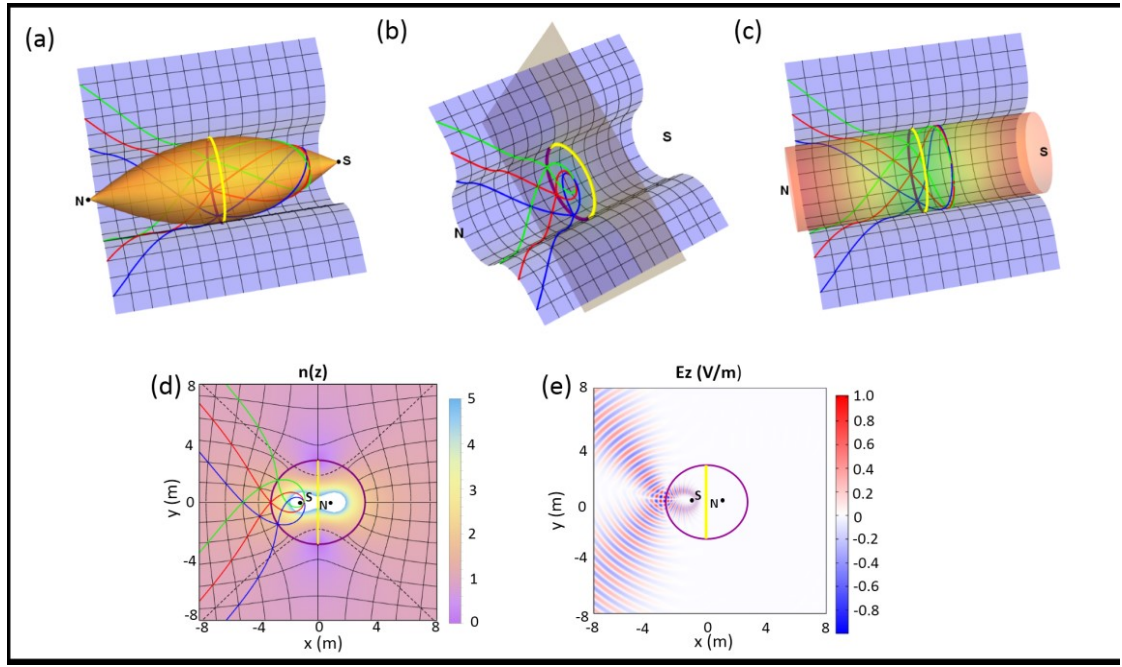

Fig. 4 Conformal reflection design with GCTO. Virtual space (a) consists of a complex plane and spindle. They are sewed along a branch cut (solid arc in purple), which belongs to geodesics of both parts, namely a straight line of plane and the equator of spindle. Using geodesic conformal mapping, spindle is mapped to a generalized Maxwell's fish-eye lens in (b). With a further exponential conformal mapping, Maxwell's fish-eye lens in (b) is mapped to a cylindrical Mikealian lens shown in (c). Using an analytical conformal mapping, physical space is finally obtained in (d), with contour plot of refractive index profile (Dashed black lines are places with refractive index of unity). Corresponding light rays are shown in colored curves in (a-d). Gaussian beam in (e) impinges our design at the angle of $-\pi/4$ with a frequency corresponding to N+m = 30. For the illustration, the length of the equator and the branch cut are set to be $4\pi$ and 12.14, respectively. The corresponding parameter *a* in an analytical conformal mapping is 1.2.

*Two methods for achieving invisibility.* – Based on the above GCTO, we can further achieve invisible cloaks with two methods. One is that we employ the fact that there is some place in virtual space, where light rays coming from the outside could not reach, such as the two yellow curves in Fig. 2(a). These two yellow curves could be equipped with PECs. This way, we can naturally achieve invisible cloaks using PECs. This method is similar to that used in non-Euclidean cloak[40]. The other method is that we can place a PEC along a meridian of geodesic lens as shown in black closed curve of Fig. 5 (a), (c) and (e) to reflect light rays twice, which will give an invisibility cloaks as well[26]. The PEC divides the branch cut into two equal parts. As shown in Fig. 5 (b), (d) and (f), it seems that these PECs in black lines are invisible, similar to that in Fig. 2(f), Fig. 3(e) and Fig. 4(d), respectively. We call this method "double-reflection mechanism". Once we make the PEC invisible, we can further expand it to make cloaking region of cloaking by another conformal mapping[45].

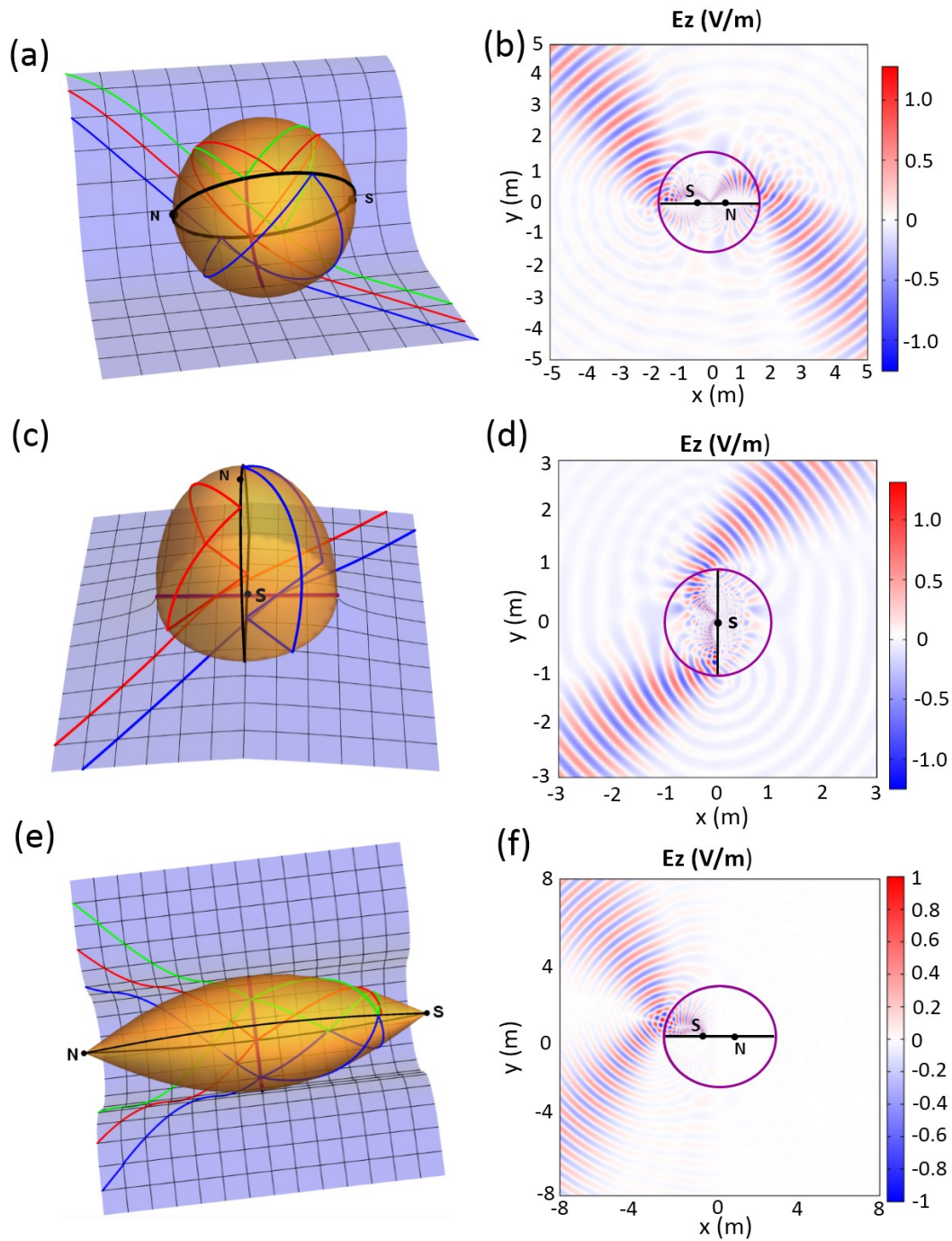

Fig. 5 Invisibility cloaks of PECs. PECs (in black) along closed geodesics of geodesic lenses of Fig. 2(a), Fig. 3(a) and Fig. 4(a), are located to divide the branch cut into equal parts. Once light rays meet the PECs, they are reflected. Light rays are shown in colored curves in (a), (c) and (e). Gaussian beams in (b), (d) and (f) are consistent with trajectories of light rays at the same condition of corresponding physical spaces of Fig. 2(e), Fig. 3(d) and Fig. 4(e).

In the first design of the above conformal transparency, we can use another PEC, which gives the double-reflection mechanism as well. One such PEC is a great circle of the sphere in black as shown in Fig. 6 (a). This PEC touches one vertex of the branch cut. After light rays enter the sphere, they will be reflected by the PEC twice and form closed trajectories. Therefore light rays will never reach the other half of the sphere because of the PEC. After mapping virtual space to physical space as in Fig. 6(b), we obtain a cloaking region bounded by PEC in black. We can also modify our device by removing a part of the PEC as is shown in Fig. 6(c-d) or, alternatively, Fig. 6(e-f). In this case the remaining PEC will reflect light rays twice and the sphere will still be invisible, but the cloaked region will be lost.

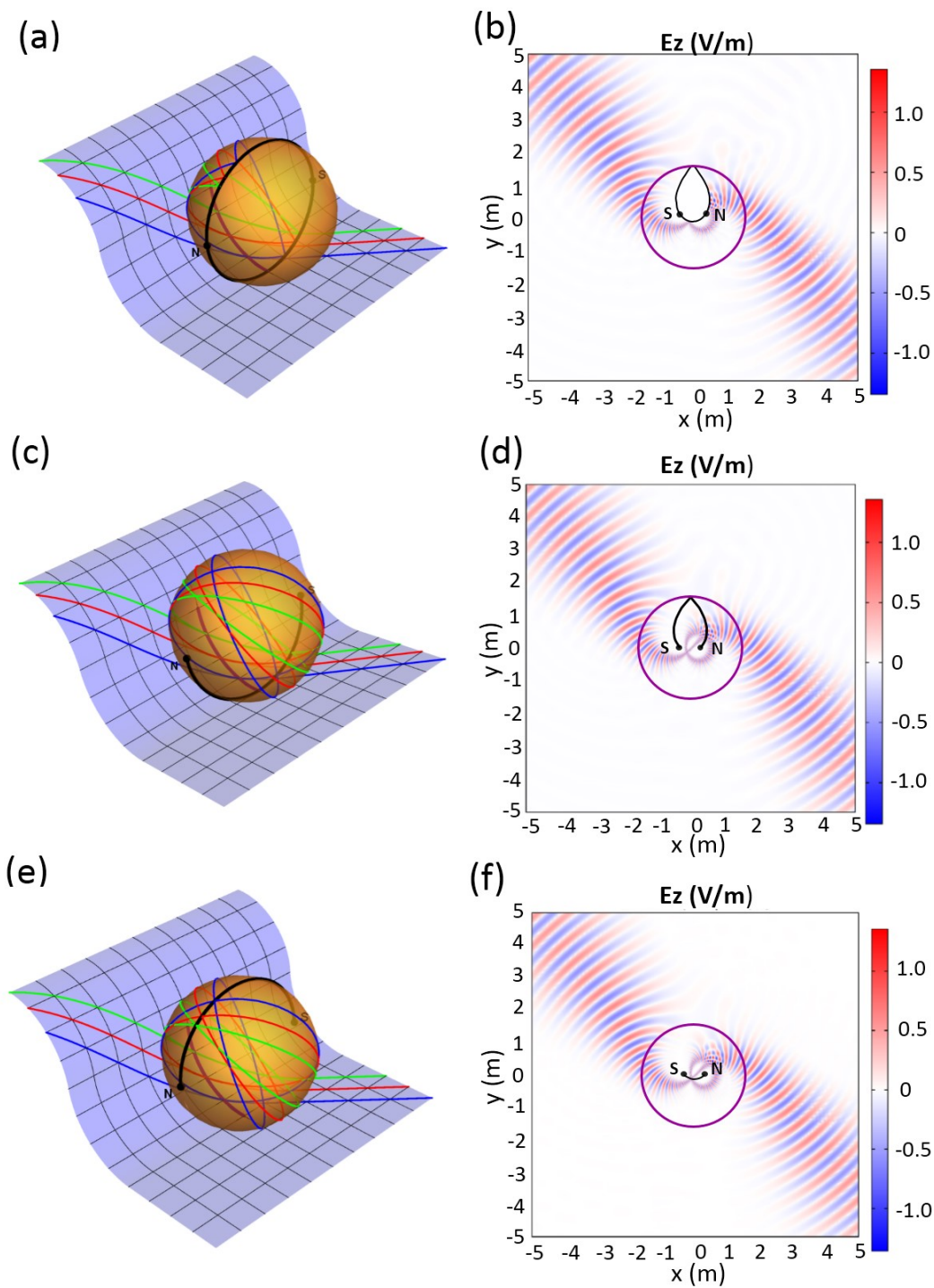

Fig. 6 Conformal cloaks of PECs. (a) PECs (in black) along closed geodesic of the sphere of Fig. 2(a) is located to pass through one vertex of the branch cut, N and S poles. Double reflection on this PEC could make half of the sphere invisible. (b) This way, a waterdrop-shape cloaking region is obtained in physical space. In fact, both halves of this PEC, which connects N and S pole, are invisible by double-reflection mechanism as shown in (c) and (e), respectively. Light

rays are shown in colored curves in (a), (c) and (e). Propagation of a Gaussian beam with the same parameters as in Fig. 2(e), is shown in (b,d,f).

In conclusion, we propose geodesic conformal transformation optics (GCTO) based on the concept of geodesic lenses. The resulting refractive index profile is continuous and shows almost perfect performance of the designed devices. We demonstrate our method by achieving optical transparency and wave reflection. Furthermore, we can achieve invisible cloak with two methods. Thanks to the intuition of virtual space, we further explain the mechanism of conformal transparency at eigen-frequencies in wave optics. In one word, we can manipulate light propagation with continuous scalar field of refractive index profile, which could be treated as the scalar curvature of two-dimensional space. The concept of GCTO may also find applications in other waves, such as acoustics, which obey the Helmholtz equation in two dimensions.

Acknowledgements

National Science Foundation of China for Excellent Young Scientists (grant no. 61322504); Foundation for the Author of National Excellent Doctoral Dissertation of China (grant no. 201217); Fundamental Research Funds for the Central Universities (Grant No. 20720170015). L. X. was supported by the China Scholarship Council for half-year study at Masaryk University. T. T. was supported by Grant No. P201/12/G028 of the Czech Science Foundation.